\documentclass[11pt]{article}
\usepackage{mathrsfs}
\usepackage{amsfonts}
\usepackage{amssymb}
\usepackage{bbm}
\usepackage{color}
\usepackage{amsfonts,amssymb, mathrsfs, amsmath,   amssymb,  theorem,  float}
\usepackage{graphicx}  

\usepackage{geometry}
\newtheorem{theorem}{Theorem}[section]

\newtheorem{cor}[theorem]{Corollary}
\newtheorem{lemma}[theorem]{Lemma}
\newtheorem{alg}[theorem]{Algorithm}
\newtheorem{remark}[theorem]{Remark}

\def\qed{\hfil {\vrule height5pt width2pt depth2pt}}

\def\qed{\hfil {\vrule height5pt width2pt depth2pt}}
\def\bref#1{(\ref{#1})}

\def\qed{\hfil {\vrule height5pt width2pt depth2pt}}

\def\C{\mathcal{C}}

\def\proof{{\noindent\em Proof.\,\,}}

\def\SL{{\mathbf{SLin}}}

\def\bref#1{(\ref{#1})}

\def\N{{\mathbb N}}
\def\Z{{\mathbb Z}}
\def\Q{{\mathbb Q}}

\def\C{{\mathbb C}}

\def\X{{\mathbb{X}}}

\def\Pr{{\mathbf{P}}}
\def\EV{{\mathbf{E}}}

\def\F{{\mathcal {F}}}

\def\RB{{\mathcal{R}}}

\def\bref#1{(\ref{#1})}

\def\+{ \oplus}
\def\-{\ominus}
\def\*{\otimes}

\def\deg{\hbox{\rm{deg}}}

\def\p{{\bf{p}}}

\def\modp{{\mathbf{mod}}}

\begin{document}
%
\title{Revisit Sparse Polynomial Interpolation based on\\ Randomized Kronecker Substitution\thanks{\quad Partially
       supported by a grant from NSFC (11688101).}}
\author{Qiao-Long Huang$^{1,2}$ and Xiao-Shan Gao$^{1,2}$ \\
 $^{1}$KLMM, Academy of Mathematics and Systems Science\\
 Chinese Academy of Sciences, Beijing 100190, China\\
 $^{2}$University of Chinese Academy of Sciences, Beijing 100049, China}
\date{}

\maketitle

\begin{abstract}
\noindent
In this paper, a new reduction based interpolation algorithm for black-box multivariate polynomials over finite fields is given. The method
is based on two main ingredients.
A new Monte Carlo method is given to reduce
black-box multivariate polynomial interpolation to black-box univariate polynomial interpolation over any ring.
The reduction algorithm leads to multivariate interpolation algorithms with better or the same complexities most cases when combining with various univariate interpolation algorithms.
We also propose a modified univariate Ben-or and Tiwarri algorithm over the finite field, which has better total complexity than the Lagrange interpolation algorithm.
Combining our reduction method and the modified univariate Ben-or and Tiwarri algorithm, we give a Monte Carlo multivariate interpolation algorithm, which has better total complexity in most cases for sparse interpolation of black-box
polynomial over finite fields.

\vskip10pt
\noindent{\bf Keywords}. Randomized Kronecker substitution, sparse polynomial interpolation, black-box,  Ben-or and Tiwari algorithm, finite field,  Monte Carlo algorithm.
\end{abstract}


%


\section{Introduction}

The interpolation for a sparse multivariate polynomial
 $$f=c_1m_1+c_2m_2+\cdots+c_t m_t\in\RB[x_1,\ldots,x_n]$$
given as a black-box is a basic computational problem, where $\RB$ is a ring. Here, the challenge is that both the monomials $m_i$ and the coefficients $c_i$ are unknown and the algorithm also needs to take advantage of the sparse structure of $f$.
%

In \cite{191}, Zippel gave a probabilistic algorithm which needs an
upper bound for the number of terms of $f$ and an upper bound for
the degree of $f$ in each variable.
In \cite{12}, Ben-Or and Tiwari gave a deterministic algorithm over the field of complex numbers,  which needs an upper bound of the number of terms in $f$.
After these work, many interesting algorithms  were given, such as
the computational complexity enhancement \cite{7,19},
the interpolation with nonstandard bases \cite{22},
the interpolation over finite fields \cite{4,6,14,16,21},
the early termination algorithm \cite{kal-e1,ka-e2},
the hybrid interpolation algorithm \cite{15,8,lee-new,15},
the interpolation for modular black-box polynomials \cite{BGbbi},
and the reduction based methods for black-box and SLP polynomials \cite{2,3,1,20,14,huang-slp}.

The sparse interpolation algorithms can be roughly divided into
two types according to the way of doing interpolation:
(1) the direct methods, such as the Ben-Or and Tiwari algorithm,
which find the monomials $m_i$ directly and then find the coefficients;
(2) the reduction methods, such as Zippel's algorithm, which
reduce the multivariate interpolation into the univariate interpolation.
Each type has its advantages and disadvantage.

The size of an $n$-variate polynomial $f$
with a degree bound $D$ and a term bound $T$ is $O(nT\log D + T \log c)$, where $c = \max_{i=1}^t |c_i|$.
The sparse interpolation algorithms can also be roughly divided into
two types according to the complexity in $D$:
(1) the polynomial-time algorithm whose complexity is polynomial in $\log D$;
(2) the exponential algorithm whose complexity is polynomial in $D$.

Since the value of a polynomial of degree $D$ at any point other than $0,\pm 1$ will have $D$ bits or more, any algorithm whose complexity is proportional to $\log D$ cannot perform such an evaluation over $\Q$ or $\Z$.
Even for polynomials over the general finite field $\F_q$, there is no polynomial-time interpolation algorithms for the standard black-box model.
On the other hand, polynomial-time algorithms do exist for three special models.

The first model is the  precision accuracy black-box model \cite{AM95,lee-new,M95,15}, which allows for evaluations on the unit circle in some representation of a subfield of $\C$ or returns only a limited number of bits of precision for an evaluation.

The second model is the modular black-box model \cite{GR10,BJ14}, which works for the polynomials in $\Q[x_1,\dots,x_n]$.
Given a prime $p$ and an element $\theta$ in $\Z_p$, the model computes $f(\theta)$ over the field $\Z_p$. The cost of the evaluation depends on the size of $p$.

The third model is the straight-line program model \cite{2,3,10,20,14,huang-slp}, which uses the arithmetic operations in the $\RB[x_1,\dots,x_n]$ to replace the black-box evaluation.

In this paper, we focus on reduction methods for general black-box models.
Our main contribution is to give a new Monte Carlo reduction method for black-box polynomials, which leads to multivariate interpolation algorithms with better or the same complexities in most cases comparing to existing reduction method.
We also propose a modified univariate Ben-or and Tiwarri algorithm over the finite
field $\F_q$ costing $O^\thicksim(D\log q + T B) $ bit operations, where $B$ is the cost of query the black-box.
Note that the Lagrange interpolation algorithm costs
$O^\thicksim(D\log q + D B)$ bit operation, which is worse since $T\le D$.
Let $f$ be an $n$-variate polynomial with a degree bound $D$ and a term bound $T$.
Combining our reduction method and the modified univariate Ben-or and Tiwarri algorithm,
we give a multivariate interpolation algorithm whose bit complexity
is  $O^\thicksim(nTD\log q+ nT B_f)$, where $B_f$ is the cost of evaluating the black-box
that gives $f$.

\subsection{Comparing with other reduction methods}

The reduction depends on the following Kronecker type substitutions:
\begin{eqnarray}
f(x^\mathbf{s}) &=& f(x^{s_1},x^{s_2},\dots,x^{s_n})\label{eq-r1}\\
%
%
f(x^{\mathbf{s}+p\mathbf{I}_k}) &=&
f(x^{s_1},x^{s_2},\dots,x^{s_k+p},\dots,x^{s_n})\label{eq-r2}
\end{eqnarray}
where $p$ is a prime and $\mathbf{s}=(s_1, \dots,s_n)\in\N^n$
is a vector of random integers.
The substitution \bref{eq-r1} introduced in \cite{1} is called {\em randomized Kronecker substitution}.
\bref{eq-r2} was introduced in \cite{0}.

Our method builds on the work \cite{0,1}. To compare with \cite{0,1}, we first explain how these algorithms work.
The algorithm in \cite{1} has three main steps. 1. Randomly choose $O(n+\log T)$ substitutions $\mathbf{s}_i$. 2: Find a diversifying set of terms of $f$ such that a term has the same coefficient after all substitutions. 3: For each term, solve a linear system to obtain its exponents.
The algorithm in \cite{0} also has three main steps. 1: Randomly choose $\log (T)$ substitutions $\mathbf{s}_i$. 2: Find  the $f(x^{\mathbf{s}_u})$ with the maximal number of terms.  3: Find a prime $p$ such that $\#f(x^{\mathbf{s}_u}) \modp (x^p-1)=\#f(x^{\mathbf{s}_u})$ and half of the terms of $f$ can be recovered from $f(x^{\mathbf{s}_u})$ and $f(x^{\mathbf{s}_u+p\mathbf{I}_k}),k=1,2,\dots,n$.

Our algorithm works as follows. 1: Randomly choose $\log (T)$ primes $p_i$ of size $O^\thicksim(T\log D)$ and substitutions $\mathbf{s}_i\in \Z^n_{p_i}$.
2: Find a $u$ such that $\#f(x^{\mathbf{s}_u})\ \modp \ (x^{p_u}-1)$ has the maximal number of terms.  3: Half of the terms of $f$ can be recovered from $f(x^{\mathbf{s}_u})$ and $f(x^{\mathbf{s}_u+p_u\mathbf{I}_k}),k=1,2,\dots,n$.

Our method is different from that in \cite{0,1} in the following aspects.
Comparing to  \cite{1},  we do not need to solve linear systems, so our algorithm is linear in $n$ while theirs is linear in $n^{\omega}$.
Also, our algorithm does not need to find the diversifying set, so it works for more general rings.
%
%
Comparing to  \cite{0}, our algorithm chooses a prime $p_i$ first and then chooses the substitutions $\mathbf{s}_i\in\Z_{p_i}^n$, while in \cite{0}, the prime is fixed.
As a consequence, the univariate polynomials in our algorithm
have degrees $O^\thicksim(TD)$, while the degrees of the univariate polynomials in \cite{0} contain either $T^2$ or $D^2$.
%

In Table \ref{tab-2}, we list the complexities of the reduction methods, where ``$\#$Reductions(N)" is the number of univariate interpolations, ``Degree" is the degree bound of the univariate polynomials, ``Extra bit complexity($\eta$)" is the additional complexities
needed besides the univariate interpolations.
``Type" means whether the algorithm is deterministic (Det),
Monte Carlo (MC), or Las Vegas (LV).
$\overline{D}=\max_{i=1}^n\deg_{x_i}(f)$.

\noindent\begin{table}[H]\footnotesize
\centering
 \scalebox{0.88}[0.80]{
\begin{tabular}{c|c|c|c|c}
%
&$\#$Reductions(N)&Degree $(\widetilde{D})$ & Extra bit cost($\eta$)&Type \\\cline{1-5}
Kronecker&1&$D^n$&$n^2T\log D$&Det\\\cline{1-5}
Zippel~\cite{191}&$nT$&$D$&$\geq nT\log D$&MC\\\cline{1-5}
Klivans-Spielman~\cite{4}&$n$&$nT^2D$&$nTD^{O(1)}$&MC\\
\cline{1-5}
               &$n\log T$&$TD$&&\\
Arnold~\cite{0}&+&$+$&$nT\log D$&MC\\
               & $\log^2 T$&$ TD+\overline{D}T\min(D,T\log(TD))$&&\\
\cline{1-5}
Arnold-Roche~\cite{1}&$n+\log T$&$TD$&$n^2 T+nT\log D$&MC\\
\cline{1-5}
Huang and Gao~\cite{huang-bb} &$n\log T$&$nTD$&$nT\log D$&MC\\\cline{1-5}
This paper  (rem. \ref{rem-1}) &$n\log T+\log ^2T$&$TD$&$nT\log D$&MC

\end{tabular}}
\caption{Reduction of multivariate polynomial interpolations to univariate ones}\label{tab-2}
\end{table}

We now compare the complexities of multivariate interpolations using the reductions given in Table \ref{tab-2}. Two cases are considered according to the complexity of the univariate interpolation algorithm to be used.

First, assume an univariate interpolation algorithm is polynomial-time with complexity $\SL(T^{\alpha},$ $\log^{\beta} D)$, where $\SL(a,b,\dots)$ means the complexity is soft-linear in $a,b,\dots$.
Then the complexities of the multivariate interpolation is
$\SL(NT^{\alpha},N\log^{\beta} \widetilde{D}, \eta)$, where
$N$, $\widetilde{D}$, and $\eta$ are from Table \ref{tab-2}.
We list these complexities in Table \ref{tab-4}.
From the table, we can see that, for the polynomial-time algorithms, our reduction method is the same as the   method in \cite{0,huang-bb} and is better than others.

\noindent\begin{table}[H]\footnotesize
\centering
 \scalebox{1.00}[1.00]{%
\begin{tabular}{c|c|c}
&Complexity&type \\\cline{1-3}
Kronecker&$\SL(n^{\max(\beta,2)},T^{\alpha},\log^{\beta} D)$&Det\\
Zippel~\cite{191}&$\SL(n,T^{\alpha+1},\log^{\beta} D)$&MC\\
Klivans-Spielman~\cite{4}&$\SL(n,T^{\alpha},\log^{\beta} D)+nTD^{O(1)}$&MC\\
Arnold~\cite{0}&$\SL(n,T^{\alpha},\log^{\beta} D)$&MC\\
Arnold-Roche~\cite{1}&$\SL(n,T^{k_1},\log^{k_2} D, n^\omega T)$&MC\\
Huang-Gao~\cite{huang-bb} &$\SL(n,T^{\alpha},\log^{\beta} D)$&MC\\
This paper (Thm. \ref{th-ff}) &$\SL(n,T^{\alpha},\log^{\beta} D)$&MC
\end{tabular}}
\caption{Complexity for polynomial-time multivariate  interpolation algorithms}\label{tab-4}
\end{table}

Second, assume an univariate   algorithm is exponential with complexity
 $\SL(T^{\alpha},D^{\beta})$.
Then the complexities of the multivariate algorithms are $\SL(NT^{\alpha}, N(\widetilde{D})^{\beta}),\eta)$, which are listed in Table \ref{tab-5}.
From the table, we can see that, for the exponential algorithms, the complexity of our algorithm is better than all the existed Kronecker-type substitutions \cite{4,0,1,huang-bb}.
Comparing to Zippel's reduction \cite{191}, our method has better, equal, or worse complexities if
$0<\beta<1$, $\beta=1$, or $\beta>1$.

\noindent\begin{table}[H]\footnotesize
\centering
 \scalebox{1.00}[1.00]{%
\begin{tabular}{c|c|c}
&Complexity&type \\\cline{1-3}
Kronecker&$\SL(T^{\alpha},D^{n\beta})+n^2T\log D$&Det\\
Zippel~\cite{191}&$\SL(n,T^{\alpha+1},D^{\beta})$&MC\\
Klivans-Spielman~\cite{4}&$\SL(n^{\beta+1},T^{\alpha+2\beta},D^{\beta})+nTD^{O(1)}$&MC\\\cline{1-3}
&$\SL(n,T^{\alpha+\beta},D^{2\beta})$&\\
Arnold~\cite{0}&or&MC\\
&$\SL(n,T^{\alpha+2\beta},D^{\beta})$&\\\cline{1-3}
Arnold-Roche~\cite{1}&$\SL(n,T^{\alpha+\beta},D^{\beta})+n^\omega T$&MC\\
Huang-Gao~\cite{huang-bb} &$\SL(n^{\beta+1},T^{\alpha+\beta},D^{\beta})$&MC\\
This paper (Thm. \ref{th-ff}) &$\SL(n,T^{\alpha+\beta},D^{\beta})$&MC
\end{tabular}}
\caption{Complexity for exponential multivariate interpolation algorithms}\label{tab-5}
\end{table}

Table \ref{tab-3} is a summary of the comparisons, where $``\surd"$, $``="$, $``\times"$ means that our reduction method has better, the same, and  worse complexity, respectively.
We can see that, for $0<\beta<1$, our reduction method is the achieve the best complexity,
and the only case our reduction has worse complexity is
for exponential algorithms with $\beta>1$.

\noindent\begin{table}[H]\footnotesize
\centering
 \scalebox{0.8}[0.8]{
\begin{tabular}{c|c|c|c|c|c|c|c}
%
\multicolumn{2}{c|}{}&Kronecker&Zippel~\cite{191}& Klivans-Spielman~\cite{4}&Arnold~\cite{0}&Arnold-Roche~\cite{1}& Huang-Gao(MC)~\cite{huang-bb}\\\cline{1-8}
\multicolumn{2}{c|}{Polynomial-time}&$\surd$&$\surd$&$\surd$&$=$&$\surd$&$=$\\\cline{1-8}
 &$0\leq \beta<1$&$\surd$&$\surd$&$\surd$&$\surd$&$\surd$&$\surd$\\\cline{2-8}
Exponential&$\beta=1$&$\surd$&$=$&$\surd$&$\surd$&$\surd$&$\surd$\\\cline{2-8}
 &$\beta>1$&$\surd$&$\times$&$\surd$&$\surd$&$\surd$&$\surd$\\\cline{1-8}
\end{tabular}}
\caption{Compare to other reduction methods}\label{tab-3}
\end{table}

Finally, we remark that the cases $\beta>1$ and $\beta <1$ do exist.
The original Ben-or and Tiwarri algorithm works  for univariate polynomials over the finite field $\F_q$
and costs $O^\thicksim(T^{1.5}\sqrt{D}\log q+T\log^2 q)$ bit operations (Refer to Remark \ref{rem-dl2}), where $\beta = 0.5$.
%
The bit complexity of the Lagrange interpolation algorithm over $\Q$ is  $O^\thicksim(D^2)$.

\subsection{Comparing with interpolation algorithms over finite fields}

In order to obtain a reduction based multivariate interpolation
algorithm, we need univariate interpolation algorithms
with best complexities.

Let $h$ be a black-box univariate polynomial in $\F_q[x]$ with a degree bound $D$ and a term bound $T$. Let $B_h$ be the cost of query the black-box.
In this paper, we gave a modified univariate Ben-or and Tiwarri algorithm
which costs $O^\thicksim(D\log q ) $ bit operations
and $O(T)$ evaluations of $h$, so the total cost is
$O^\thicksim(D\log q + T B_h) $.
The Lagrange interpolation algorithm costs $O^\thicksim(D\log q)$ bit operations and $O(D)$ evaluations of $h$ and the total complexity is
$O(D\log q + D B_h)$. So, the modified univariate Ben-or and Tiwarri
algorithm has lower complexities than the Lagrange algorithm.

An univariate Ben-or and Tiwari algorithm over the finite filed was given in \cite{10},  whose complexity includes the parameter $q$.
Also, the multivariate Ben-or and Tiwari algorithm was extended to  finite fields \cite{16,21}, whose complexities are quite high (see Table \ref{tab-1}).

Combing the modified univariate Ben-or and Tiwarri algorithm
and our reduction method, we give a new  multivariate interpolation algorithm.
Table \ref{tab-1} is a comparison with interpolation algorithms over finite fields.
``Probes" is the number of evaluations for the polynomials,
``Bit complexity" is the complexity besides the probes,
and ``Size of $\F_q$" means that the algorithm can work for the finite field whose size satisfies this condition, and in the contrary case, the algorithm need to take values in a proper extension field of $\F_q$.

\begin{table}[H]
\centering
\scalebox{0.80}[0.80]{%
\begin{tabular}{c|c|c|c|c}
&Probes $(\rho)$ &Bit complexity $(\Theta)$ &Size of $\F_q$&type \\\cline{1-5}
Grigoriev-Karpinski-Singer~\cite{6}&&$n^2T^6\log^2(ntq)+q^{2.5}\log^2q$&&Det\\
Huang-Rao~\cite{16}&$T^2D$&$(TD)^8((TD)^5+\log q)\log^2 q+nTD\log q$&$q\geq O(T^2D^2)$&LV\\
Javadi and Monagan~\cite{21}&$nT$&$T^2(\log q+nD)\log q$&$\phi(q-1)\geq O(nD^2T^2)$&MC\\
%
Klivans-Spielman~\cite{4}&$nT$&$n^2T^2D\log q$&$q\geq O(nT^2D)$&MC\\
Arnold-Roche~\cite{1}&$nT$&$nTD\log q+n^\omega T$&$q\geq O(TD)$&MC\\
Huang-Gao~\cite{huang-bb} &$nT$&$n^2TD\log q$&$q\geq O^\thicksim(nDT)$&MC\\
Zippel~\cite{191,21} &$nTD$&$nTD\log q$&$q\geq O(nD^2T^2)$&MC\\\cline{1-5}

This paper (Thm. \ref{th-ff}) &$nT$&$nTD\log q$&$q\geq O(TD)$&MC\\
This paper (Rem. \ref{rem-dl2}) &$nT$&$nT^{1.5}\sqrt{D}\log q+nT\log^2 q$&$q\geq O(TD)$&MC
\end{tabular}}
\caption{``Soft-Oh" comparison of interpolation algorithms over finite field $\F_q$ }\label{tab-1}
\end{table}

The total complexity of an algorithm is $O^\thicksim(\Theta + \rho B )$,
where $\Theta$ and $\rho$ are from Table \ref{tab-1} and  $B $
is the cost of probing the black-box.
The bit complexities of the algorithms given in \cite{6,16} are
much higher than other algorithms, so we will not compare with them below.

We can see that our algorithm (Thm. \ref{th-ff}) has better total complexity than all other methods in \cite{21,4,1,huang-bb,191,21}.
Comparing to Zippel's algorithm, our algorithm has the same bit complexity but needs less evaluations and works for a smaller field.
Actually, our algorithm is the only one which achieves the
best current bounds in all three parameters in Table \ref{tab-1}.

The algorithm given in Remark \ref{rem-dl2} uses the original Ben-or and Tiwarri algorithm works   univariate polynomials over the finite field $\F_q$,
which costs $O^\thicksim(nT^{1.5}\sqrt{D}\log q+nT\log^2 q)$ bit operations.
By Table \ref{tab-3}, if using this univariate interpolation algorithm,
our reduction method gives the multivariate interpolation algorithm
with best complexities comparing with other reduction methods.

%
%
%

\section{Reduction   based on randomized Kronecker substitution}
In this section, we give a new Monte Carlo  algorithm which reduces multivariate polynomial interpolation to that of univariate polynomial interpolation based on randomized Kronecker substitutions over any commutative ring with identity.

\subsection{Find an ``ok" random Kronecker substitution}
Let $f\in\RB[\X]$, where $\RB$ is commutative ring with identity
and $\X=\{x_1,x_2,\dots,x_n\}$ is a set of $n$ indeterminates.
Denote $\#f$ and $\deg f$ to be the number of terms in $f$
and the total degree of $f$, respectively.
For $\mathbf{s}=(s_1,s_2,\dots,s_n)\in\N^n$ and a new indeterminate $x$, let
\begin{eqnarray}
&&f(x^\mathbf{s})=f(x^{s_1},x^{s_2}, \dots,x^{s_n})\label{eq-rk1}\\
&&f^{\modp}_{(p)}(x^\mathbf{s}) = f(x^{s_1},x^{s_2}, \dots,x^{s_n})\, \modp \ (x^p-1).\label{eq-rkm}
\end{eqnarray}
%
%
For $\mathbf{s}=(s_1,s_2,\dots,s_n)\in\N^n$,
a term $cm_1$ of $f$ is said to {\em collide} in $f(x^{\mathbf{s}})$
(or other univariate reductions of $f$) if $f$ has another term $em_2$
such that $m_1\ne m_2$ and $m_1(x^\mathbf{s})=m_2(x^\mathbf{s})$.

When $\mathbf{s}=(s_1,s_2,\dots,s_n)$ is chosen randomly,
the substitution $x_i=x^{s_i},i=1,2,\dots,n$ is called a
{\em randomized Kronecker substitution}.
For a prime $p$, a  substitution $\mathbf{s}$ is called
{\em ``ok" with respect to $p$}, if a majority, say $\frac58$, of the terms of $f$ do no collide in  $f^{\modp}_{(p)}(x^{\mathbf{s}})$.

We need the following Hoeffding's inequility for Bernoulli random variables.
\begin{lemma}\cite{5}
Let $X=\sum_{i=1}^nX_i$, where $X_i$,$i=1,2,\dots,n$, are independently distributed in $[0,1]$. Then for all $\varepsilon>0$, $\Pr[X>\EV[X]+\varepsilon]\leq e^{-2\varepsilon^2/n}$ and $\Pr[X<\EV[X]-\varepsilon]\leq e^{-2\varepsilon^2/n}$, where
$\EV[X]$ is the expected value of $X$.
\end{lemma}

We  have the following key lemma.
\begin{lemma}\label{lm-1}
Let $f=\sum_{i=1}^tc_im_i\in\RB[\X]$, $T\ge \#f$, $D\ge \deg f$, and $N=\max\{31\lfloor(T-1)\log_2 D\rfloor,1\}$. Let $p_1,p_2,\dots,p_N$ be $N$ different primes which satisfy $p_i\geq 32(T-1)$. If we randomly choose a prime $p$ in $\{p_1,p_2,\dots,p_N\}$ and  choose $\mathbf{s}\in\Z_p^n$ uniformly at random, where $\Z_p=\{0,1,\ldots,p-1\}$. Then any fixed term of $f$ collides in $f^{\modp}_{(p)}(x^\mathbf{s})$ with probability $\le \frac{1}{16}$.
\end{lemma}
\proof
If $t=1$ or $D=1$, then the proof is obvious. So now we assume $T\geq t\geq 2$ and $D\geq 2$. In this case, $N=31\lceil(T-1)\log_2 D\rceil$.
Assume $m_i=x_1^{e_{i,1}}x_2^{e_{i,2}}\cdots x_n^{e_{i,n}},i=1,2,\dots,t$. Without loss of generality, we consider the first term $c_1m_1$. Let $h(s_1,s_2,\dots,s_n)=\prod_{i=2}^t[(e_{i,1}-e_{1,1})s_1+(e_{i,2}-e_{1,2})s_2+\cdots+(e_{i,n}-e_{1,n})s_n]\ $ which
 is a polynomial in  $\Z[s_1,s_2,\dots,s_n]$ with degree no more than $T-1$. Assume the variables are ordered as $s_1\prec s_2\prec \cdots\prec s_n$ and $k_i$ is the largest number such that $e_{i,k_i}-e_{1,k_i}\neq 0$. Then $\prod_{i=2}^t[e_{i,k_i}-e_{1,k_i}]s_{k_i}$ is the leading term.

Let $C=\prod_{i=2}^t[e_{i,k_i}-e_{1,k_i}]$
and let $k$ be the number of different prime factors of $C$. Since $|e_{i,k_i}-e_{1,k_i}|\leq D$,
we have $2^k \le C$ and hence  $C$ has at most $\lfloor(T-1)\log_2 D\rfloor$ different prime factors. So if we randomly choose a prime $p$ in $\{p_1,p_2,\dots,p_N\}$, with probability at least $1-\frac{\lfloor(T-1)\log_2D\rfloor}{N}=\frac{30}{31}$, $\prod_{i=2}^t[e_{i,k_i}-e_{1,k_i}]\ \modp\ p\neq 0$. In this case, $h(s_1,\dots,s_n)$ $\modp\ p$ is a non-zero polynomial in $\F_p[s_1,s_2,\dots,s_n]$.

If $h(s_1,\dots,s_n)\ \modp\ p\neq 0$, then by Zippel's lemma \cite{191}, if we choose $\mathbf{s}\in\Z_p^n$ uniformly at random, then $h(\mathbf{s})\ \modp\ p\neq 0$ with probability at least $1-\frac{T-1}{p}\geq1-\frac{T-1}{32(T-1)}=\frac{31}{32}$.

So if we randomly choose a prime $p$ in $\{p_1,p_2,\dots,p_N\}$ and  choose $\mathbf{s}\in\Z_p^n$ uniformly at random, with probability at least $\frac{30}{31}\cdot \frac{31}{32}=\frac{15}{16}$, $h(\mathbf{s})\ \modp\ p\neq0$.

Now it suffices to show that when $h(\mathbf{s})\ \modp \ p\neq 0$,  $c_1m_1$ does not collide in $f^{\modp}_{(p)}(x^\mathbf{s})$. Since $h(\mathbf{s})\ \modp\ p \neq 0$, $(e_{i,1}-e_{1,1})s_1+(e_{i,2}-e_{1,2})s_2+\cdots+(e_{i,n}-e_{1,n})s_n\neq 0\ \modp\ p$. So $(e_{i,1}s_1+e_{i,2}s_2+\cdots+e_{i,n}s_n)\ \modp\ p\neq (e_{1,1}s_1+e_{1,2}s_2+\cdots+e_{1,n}s_n)\ \modp\ p$, which means that $c_im_i$ does not collide with $c_1m_1$ in $f^{\modp}_{(p)}(x^\mathbf{s})$. \qed

We also need the following lemma.
\begin{lemma}\label{lm-3}
Let $B_j,j=1,2,\dots,s$ be nonempty sets of integers  and $a_i,i=1,2,\dots,t$ all the different elements in $\cup_{j=1}^s B_j$. Let $c$ be the number of $a_i$ satisfying $a_i\in B_j$ and $\#B_j\geq 2$ for some $j$.
Then $t-c\le s$ and for $s_1\in[t-c,s]\cap\N$, we have $(t-s_1)\leq c\leq2(t-s_1)$.
\end{lemma}
\proof
$B_j$ is called a single point set if $\#B_j=1$,
and a collision set if $\#B_j\geq 2$.
Since $t-c$ is the number of $a_i$ contained in all single point sets, there exist $t-c$ single point sets. So $t-c\leq s$.  Since $t-c\leq s_1\leq s$, we have $(t-s_1)\leq c$.
Let $k_1$ be the number of collision sets. We have $k_1+t-c=s$. So $c=k_1+t-s$. Since every collision set contains at least two elements, $k_1\leq \frac12 c$. So $c\leq \frac12 c+t-s$, which is $\frac12 c\leq t-s\leq t-s_1$. So $c\leq2(t-s_1)$. \qed

For $p\in\Z_{>0}$ and $\mathbf{u}\in \Z^n$,
let $\mathcal{C}_f(p,\mathbf{u})$ be the number of terms of $f$ that collide in $f^{\modp}_{(p)}(x^\mathbf{u})$.
\begin{lemma}\label{lm-2}
Let $p_\mathbf{u},p_\mathbf{v}\in\Z_{>0}$ and $\mathbf{u},\mathbf{v}\in\Z^n$ such that
$\#[f^{\modp}_{(p_\mathbf{u})}(x^\mathbf{u})]\geq \#[f^{\modp}_{(p_\mathbf{v})}(x^\mathbf{v})$. Then $\mathcal{C}_f(p_\mathbf{u},\mathbf{u})\leq 2\mathcal{C}_f(p_\mathbf{v},\mathbf{v})$.
\end{lemma}
\proof Assume $\#[f^{\modp}_{(p_\mathbf{u})}(x^\mathbf{u})]=k_0,
\#[f^{\modp}_{(p_\mathbf{v})}(x^\mathbf{v})]=k$ and
$f^{\modp}_{(p_\mathbf{u})}(x^\mathbf{u})
=a_1x^{d_1}+a_2x^{d_2}+\cdots+a_{k_0}x^{d_{k_0}},d_i\neq d_j$, when $i\neq j$. Let $f=f_1+f_2+\cdots+f_{k_0}+g$, where
$(f_i)^{\modp}_{(p_\mathbf{u})}(x^\mathbf{u})=a_ix^{d_i},i=1,2,\dots,k_0$ and $g^{\modp}_{(p_\mathbf{u})}(x^\mathbf{u})=0$. Let $B_i,i=1,2,\dots,k_0$ be the set of terms in $f_i$ and $B_0$ be the set of terms in $g$.
So by Lemma \ref{lm-3}, we have $(t-k_0)< \mathcal{C}_f(p_\mathbf{u},\mathbf{u})\leq2(t-k_0)$. By the same reason, we have $(t-k)\leq\mathcal{C}_f(p_\mathbf{v},\mathbf{v})\leq2(t-k)$.
Now $\mathcal{C}_f(p_\mathbf{u},\mathbf{u})\leq 2(t-k_0)\leq2(t-k)\leq 2\mathcal{C}_f(p_\mathbf{v},\mathbf{v})$. \qed

The following theorem is similar to \cite[Prop.5.4.2]{0}
and has two differences. (1). For each substitution, we   choose a random prime, while in \cite[Prop.5.4.2]{0}, the prime is fixed. (2). We choose the substitution $\mathbf{s}$ such that $\#f^{\modp}_{(p)}(x^{\mathbf{s}})$ has the maximal number of terms, while in \cite[Prop.5.4.2]{0}, they choose the one such that $\#f(x^{\mathbf{s}})$ has the maximal number of terms.
\begin{theorem}\label{the-1}
Let $f(\X)\in\RB[\X]$, $T\geq\#f$, $D\geq\deg f$, $N=\max\{31\lceil(T-1)\log_2 D\rceil,1\}$ and $p_1,p_2,\dots,p_N$ be $N$ different primes which satisfy $p_i\geq 32(T-1)$.
Let $\mu\in(0,1)$ and $l\geq \lceil32\ln(T\mu^{-1})\rceil$. For $i=1,2,\dots,l$, we randomly choose a prime $p_{\alpha_i}$ in $\{p_1,p_2,\dots,p_N\}$ and then choose $\mathbf{s}_i\in\Z_{p_{\alpha_i}}^n$ uniformly at random.  Let $(p,\mathbf{s})$ be the vector in $\{(p_{\alpha_1},\mathbf{s}_1),(p_{\alpha_2},\mathbf{s}_2),\dots,(p_{\alpha_l},\mathbf{s}_l)\}$ such that $\#[f^{\modp}_{(p)}(x^\mathbf{s})]=
\max_{i=1}^l\#[f^{\modp}_{(p_{\alpha_i})}(x^{\mathbf{s}_i})]$. Then at least $\frac{5}{8}\#f$ terms  of $f$ do not collide in
$f^{\modp}_{(p)}(x^{\mathbf{s}})$  with probability at least $1-\mu$.
\end{theorem}
\proof
First we consider a fixed term $c_im_i$ and let
$f_j(x)= f^{\modp}_{(p_{\alpha_j})}(x^{\mathbf{s}_j})$.
By Lemma \ref{lm-1}, the probability of $c_im_i$ colliding in $f_j(x)$ is no more than $\frac{1}{16}$.
We define $X_j=1$ to be the event that $c_im_i$ collides in $f_j(x)$ and $X_j=0$ to be the event that $c_im_i$ does not collide in $f_j(x)$ for some $j$.

Define  $X=\sum_{j=1}^lX_j$, then $\EV[X]\le \frac{1}{16}l$.
By Hoeffiding's inequality, we have $\Pr(X>\frac{1}{16}l+\varepsilon)\leq \Pr(X>\EV[X]+\varepsilon)\leq e^{-2\varepsilon^2/l}$.
Let $\varepsilon=\frac{1}{8}l$, then $\Pr(X>\frac{3}{16}l)\leq
e^{-l/32}\le e^{\ln(\mu/T)} = \frac{\mu}{T}$.
So at probability  $\le1-\mu$, for all term $c_im_i$ of $f$, $c_im_i$ collides in at most $\frac{3}{16}l$ of $f_j(x), j=1,2,\dots,l$.
In other words, with probability $\ge 1-\mu$,  at leat
$\frac{13}{16}l\#f$ terms in
$f^{\modp}_{(p_{\alpha_j})}(x^{\mathbf{s}_j}), j=1,2,\dots,l$ do not collide.

We claim that at least one of  $f_j(x)$ has at least $\frac{13}{16}\#f$ non-colliding terms. We prove the claim by contradiction.
Assume that each $f_j(x)$ has $<\frac{13}{16}\#f$ non-colliding terms. Then there exist  $<\frac{13}{16}\#fl$ non-colliding terms in $f_j(x), j=1,2,\dots,l$, which contradicts to the fact that these $f_j(x)$ have $\geq\frac{13}{16}l\#f$ non-colliding terms.

So there must exist one $(p_{\alpha_j},\mathbf{s}_j)$ for which  at most $\frac{3}{16}$ of the terms of $f$ collide. By Lemma \ref{lm-2}, the polynomial with maximum $\#f_j(x)$ has $\frac{5}{8}\#f$ non-colliding terms.\qed

\subsection{Recover non-colliding terms}

For $\mathbf{s}=(s_1,s_2,\dots,s_n)\in\N^n$ and $q\in\N_{>0}$, let
\begin{eqnarray}\label{eq-rk2}
f(x^{\mathbf{s}+p\mathbf{I}_k})=f(x^{s_1},\dots,x^{s_k+q},\dots,x^{s_n})
\end{eqnarray}
to be the univariate polynomial obtained with the substitution: $x_i=x^{s_i},i=1,2,\dots,n,i\neq k,x_k=x^{s_k+q}$, where $\mathbf{I}_k\in\Z_{\geq0}^n$ is the $k$-th unit vector.

In this section, we show how to recover the non-colliding terms
of $f\in\RB[\X]$ from $f^{\modp}_{(p)}{(x^\mathbf{s})}$, $f(x^\mathbf{s})$, and $f(x^{\mathbf{s}+p\mathbf{I}_k})$.
Let  \begin{eqnarray}\label{eq-t6}
 &&f^{\modp}_{(p)}{(x^\mathbf{s})}=a_1x^{d_1}+\cdots+a_r x^{d_r} \label{eq-mfdp}
\end{eqnarray}
Since $f^{\modp}_{(p)}{(x^\mathbf{s})} = f^{\modp}_{(p)}{(x^{\mathbf{s}+p\mathbf{I}_k})}$, for $k=1,2,\dots,n$,   we can write
  \begin{eqnarray}\label{eq-t7}
 &&f{(x^\mathbf{s})}=f_1+f_2+\cdots+f_r+g\label{eq-mfp}\\
 &&f{(x^{\mathbf{s}+p\mathbf{I}_k})}=f_{k,1}+f_{k,2}+\cdots+f_{k,r}+g_k\nonumber
\end{eqnarray}
where
$f_i\ \modp\ (x^p-1)=f_{k,i}\ \modp\ (x^p-1)=a_ix^{d_i}$, $g\ \modp \ (x^p-1)=g_k\ \modp \ (x^p-1)=0$.
We define the following key notation
\begin{eqnarray}
&&TS_{(f,p,\mathbf{s},D)} =\{a_i x_1^{e_{i,1}}\cdots x_n^{e_{i,n}}| a_i \hbox{ is from } \bref{eq-mfdp}, \hbox{ and }  \cr
 &&\quad \hbox{T1}: f_i=a_ix^{u_i},f_{k,i}=a_ix^{b_{k,i}},k=1,2,\dots,n.\label{eq-uf2}\\ 
 &&\quad \hbox{T2}: e_{i,k}=\frac{b_{k,i}-u_i}{p}\in\N,k= 1,2,\dots,n.\cr
 &&\quad \hbox{T3}: u_i =e_{i,1}s_1+e_{i,2}s_2+\cdots+e_{i,n}s_n
 .\cr
 &&\quad \hbox{T4}:  \sum_{j=1}^n e_{i,j}\leq D. \}\nonumber
\end{eqnarray}

\begin{lemma}\label{lm-5}
Let $f=\sum_{i=1}^tc_im_i\in\RB[\X]$ and $D\geq\deg(f)$. If $cm$ does not collide in $f^{\modp}_{(p)}(x^{\mathbf{s}})$, then $cm\in TS_{(f,p,\mathbf{s},D)}$.
\end{lemma}
\proof
It suffices to show that $cm$ satisfies the conditions of the definition of $TS_{(f,p,\mathbf{s})}$. Assume $m=x_1^{e_1}x_2^{e_2}\cdots x_n^{e_n}$.
Since $cm$ is not a collision in $f^{\modp}_{(p)}(x^\mathbf{s})$,
without loss of generality, assume $cm(x^\mathbf{s})\ \modp\ (x^p-1) = a_1x^{d_1}$, where $a_1x^{d_1}$ is defined in \bref{eq-mfdp}.
It is easy to show that
$cm$ is also not a collision in $f(x^\mathbf{s})$ and in $f(x^{\mathbf{s}+p\mathbf{I}_k})$.
Hence, $f_{1} = a_1 x^{u_{1}}$ for $u_{1}=\sum_{i=1}^n e_is_i$;
$b_{k,1} = u_1 + pe_k$.
Clearly, T1, T2 and T3 are correct. Since $\deg(m)=\sum_{j=1}^n e_{i,j}\leq D$, T4 is correct.
\qed

Now we give the algorithm to compute $TS_{(f,p,\mathbf{s})}$.
\begin{alg}[TSTerms]\label{alg-1}
\end{alg}

{\noindent\bf Input:}

$\bullet$ Univariate polynomials $f^{\modp}_{(p)}(x^\mathbf{s}),f(x^\mathbf{s}),f(x^{\mathbf{s}+p\mathbf{I}_k})$, where $k=1,2,\dots,n$.

$\bullet$ A prime $p$.

$\bullet$ A vector $\mathbf{s}=(s_1,s_2,\dots,s_n)\in \Z^n_{\geq 0}$.

$\bullet$ Degree bound $D\geq \deg(f)$.

{\noindent\bf Output:} $\hbox{TS}_{(f,p,\mathbf{s},D)}$.

\begin{description}
\item[Step 1:]
Write $f^{\modp}_{(p)}(x^\mathbf{s})$, $f(x^\mathbf{s})$, and $f(x^{\mathbf{s}+p\mathbf{I}_k})$ in the following form
%
\begin{eqnarray*}
f^{\modp}_{(p)}(x^\mathbf{s})&=&a_1x^{d_1}+a_2x^{d_2}+\cdots+a_rx^{d_r}\cr
f(x^\mathbf{s})&=&a_1x^{u_1}+\cdots+a_{\gamma}x^{u_{\gamma}}+f_1\\
f(x^{\mathbf{s}+p\mathbf{I}_k})&=&a_1x^{b_{k,1}}+\cdots+a_{\gamma}x^{b_{k,\gamma}}+
f_{k,2}
\end{eqnarray*}

where $i=1,2,\dots,\gamma$, $k=1,\dots,n$, $a_ix^{u_i}$, $a_ix^{b_{k,i}}$ are all the terms satisfying:  $x^{b_{k,i}}$ is the unique term in $f(x^{\mathbf{s}+p\mathbf{I}_k})$ such that $\modp(b_{k,i},p)=d_i$  and  $x^{u_i}$ is the unique term in $f(x^\mathbf{s})$ such that $\modp(u_i,p)=d_i$.

\item[Step 2:] Let $S=\{\}$.

\item[Step 3:] For $i=1,2,\dots,\gamma$
    \begin{description}
    \item[a:]
    for $k=1,2,\dots,n$ do

 let $e_{i,k}=\frac{b_{k,i}-u_i}{p}$.
 If $e_{i,k}\notin \N$, then break.

    \item[b:] if $u_i\neq e_{i,1}s_1+e_{i,2}s_2+\cdots+e_{i,n}s_n$, then break;

    \item[c:] if $\sum_{j=1}^ne_{i,j}> D$, then break;

    \item[d:]  Let $S=S\bigcup\{a_ix_1^{e_{i,1}}\cdots x_n^{e_{i,n}}\}$.
    \end{description}

\item[Step 4] Return $S$.
\end{description}

\begin{lemma}\label{the-3}
Algorithm \ref{alg-1} needs $O(nT)$ ring operations in $\RB$ and $O^\thicksim(nT\log(s_{\max} D+pD))$ bit operations, where $s_{\max}=\max\{s_1,s_2,\dots,s_n\}$.
\end{lemma}
\proof
In Step 1, in order to match the terms of $f^{\modp}_{(p)}(x^\mathbf{s})$, $f(x^\mathbf{s})$, and $f(x^{\mathbf{s}+p\mathbf{I}_k})$, it needs  $O^\thicksim(nT\log(s_{\max}D+pD))$ bit operations and $O(nT)$ ring operations in $\RB$.
In Step 3, $\mathbf{a}$, $\mathbf{b}$  and $\mathbf{c}$ need $O(nT)$ arithmetic operations in $\Z$. Since the height of the data is $O(s_{\max}D+pD)$,  the complexity of Step 3 is $O(nT\log (s_{\max}D+pD))$ bit operations.\qed

\subsection{Algorithms}
We will give the reduction algorithm for $f\in\RB[\X]$, which works as follows.
We first find an ``ok" random Kronecker substitution $\mathbf{s}$ based on
Theorem \ref{the-1}, then obtain half of the terms of $f$ by applying Algorithm \ref{alg-1}, and finally repeat the procedure for at most $\log (\#f)$ times
to find $f$.
We assume an interpolation algorithm for univariate polynomials is given in advance.

We first give an   algorithm to obtain the polynomials $g(x^{\mathbf{s}+p\mathbf{I}_k}),k=1,\dots,n$ from  $g(\X)$.
%
\begin{alg}[PolySubs]\label{alg-mulp2}
\end{alg}

{\noindent\bf Input:}

$\bullet$ A polynomial  $g\in\RB[\X]$.

$\bullet$ A vector $\mathbf{s}=(s_1,s_2,\dots,s_n)\in \Z^n_{\geq 0}$.

$\bullet$ A prime $p$.

{\noindent\bf Output:} $g(x^{\mathbf{s}+p\mathbf{I}_k}),k=1,2,\dots,n$.

\begin{description}
\item[Step 1:] Assume $g=c_1m_1+c_2m_2+\cdots+c_tm_t$, where $m_i=x_1^{e_{i,1}}x_2^{e_{i,2}}\cdots x_n^{e_{i,n}},i=1,2,\dots,t$.

\item[Step 2:] 
For $i=1,2,\dots,n$,
let $h_i=0$;

\item[Step 3:]
For $i=1,2,\dots,t$ do

\begin{description}
\item[a:] Let $d=0$.
\item[b:] For $k=1,2,\dots, n$,
let $d=d+e_{i,k}s_k$.

\item[c:] For $k=1,2,\dots,n$, let
 $h_k:=h_k+c_ix^{d+e_{i,k}p}$.
\end{description}

\item[Step 4:]
Return $h_i,i=1,2,\dots,n$;

\end{description}

\begin{lemma}\label{lm-m1}
The complexity of Algorithm \ref{alg-mulp2}  is $O^\thicksim(nt\log (p+s_{\max})+nt\log(\deg (f)))$ bit operations and   $O(nt)$ arithmetic operations in $\RB$, where $s_{\max}=\max\{s_1,s_2,\dots,s_n\}$.
\end{lemma}
\proof
 In $\mathbf{b}$ of Step 3, $d$ is the degree of $m_i(x^\mathbf{s})$. In $\mathbf{c}$, since $\deg(m_i(x^{\mathbf{s}+p\mathrm{I}_k}))=\deg (m_i(x^\mathbf{s}))+pe_{i,k}$, $h_k$ is $f(x^{\mathbf{s}+p\mathbf{I}_k})$ after finishing Step 3. So the correctness is proved.

Now we analyse the complexity.
In $\mathbf{b}$ of Step 3, it needs $O(nt)$ arithmetic operations in $\Z$. Since $\deg (m_i(x^\mathbf{\mathbf{s}}))$ is $O(s_{\max}\deg(f))$, the bit operation is $O(nt\log(s_{\max}\cdot\deg (f)))$.

In $\mathbf{c}$, it needs $O^\thicksim(nt\log (p\cdot\deg(f)+s_{\max}\cdot\deg (f)))$ bit operations and at most $O(nt)$ arithmetic operations in $\RB$.\qed

Now we give an algorithm which interpolates at least half of the terms.

\begin{alg}[HalfPoly]\label{alg-2}
\end{alg}

{\noindent\bf Input:}

$\bullet$ A black-box procedure $\mathcal{B}_f$ that computes $f\in\RB[x_1,\dots,x_n]$.

$\bullet$ A polynomial $f^*\in\RB[x_1,\dots,x_n]$.

$\bullet$ Term bounds $T\geq \max(\#f,\#f_1),T_1\geq \#(f-f^*)$ and $T\geq T_1$.

$\bullet$ Degree bound $D\geq \max(\deg(f),\deg (f^*))$.

$\bullet$ A tolerance $\nu$ such that $0<\nu<1$.

{\noindent\bf Output:}
With probability $\geq1-\nu$, return a polynomial $h$ such that $\#(f-f^*-h)\leq \lfloor\frac{T_1}{2}\rfloor$.

\begin{description}
\item[Step 1:] Let $l= \lceil32\ln(T_1\nu^{-1})\rceil,N=\max\{31\lfloor(T_1-1)\log_2 D\rfloor,1\}$. Find the first $N$ primes $\{p_1,p_2,$ $\dots,p_N\}$ such that $p_i\geq32(T_1-1)$.

\item[Step 2:] For $i=1,\dots,l$, randomly choose $p_{\alpha_i}$ in $\{p_1,\dots,p_N\}$, then choose $\mathbf{s}_i\in\Z_{p_{\alpha_i}}^n$ uniformly at random.
    Deleting the repeated numbers, we still denote these vectors as $(p_{\alpha_1},\mathbf{s}_1),(p_{\alpha_2},\mathbf{s}_2),$ $\dots,(p_{\alpha_l},\mathbf{s}_l)$.

\item[Step 3:] For $i=1,2,\dots,l$, compute $f(x^{\mathbf{s}_i})$ from $\mathcal{B}_f$ by a given univariate interpolation algorithm with degree bound $\|\mathbf{s}_i\|_\infty D$ and term bound $T$. Let $f_i=f(x^{\mathbf{s}_i})-f^*(x^{\mathbf{s}_i})$ and $f_i^{\modp}=f_i\ \modp\ (x^{p_{\alpha_i}}-1)$.

\item[Step 4:]
Find $j_0$ such that $\#f^{\modp}_{j_0}=\max\{\#f^{\modp}_i|i=1,2,\dots,l\}$.
    If $\#f^{\modp}_{j_0}\geq T_1$, return failure.

\item[Step 5:] For $k=1,2,\dots,n$, find $f(x^{\mathbf{s}_{j_0}+p_{\alpha_{j_0}}\mathbf{I}_k})$ from $\mathcal{B}_f$ by the given univariate interpolation algorithm with degree bound $\|\mathbf{s}_{j_0}+p_{\alpha_{j_0}}\mathbf{I}_k\|_\infty D$ and term bound $T$.
Let $\{f^*_1,f^*_2,\dots,f^*_n\}=\mathbf{PolySubs}(f^*,\mathbf{s}_{j_0},p_{\alpha_{j_0}})$.
    Let $g_k=f(x^{\mathbf{s}_{j_0}+p_{\alpha_{j_0}}\mathbf{I}_k})-f^*_k$.

\item[Step 6:]
Let
$\hbox{TS}=\mathbf{TSTerms}(f^{\modp}_{j_0},f_{j_0},g_1,g_2,\dots,g_n,p_{\alpha_{j_0}},\mathbf{s}_{j_0},D)$.

\item[Step 7:]

Return $h=\sum_{s\in \hbox{TS}} s$.
\end{description}

\begin{lemma}\label{them-3}
Algorithm \ref{alg-2} computes $h$ such that $\#(f-f^*-h)\leq \lfloor\frac{T_1}{2}\rfloor$ with probability $\ge 1-\nu$.
The algorithm needs

$\bullet$ $O(n+\log T+\log \frac{1}{\nu})$ interpolations of univariate polynomials of degree $O^\thicksim(TD)$ and sparseness $\le T$.

$\bullet$ $O^\thicksim(nT\log \frac{1}{\nu})$ additional ring operations  and $O^\thicksim(nT\log D\log\frac{1}{\nu})$ additional bit operations.
\end{lemma}
\proof
We first show that Algorithm \ref{alg-2} returns the polynomial $h$ such that $\#(f-f^*-h)\leq \lfloor\frac{T_1}{2}\rfloor$ with probability $1-\nu$.
In Step 1 and Step 2, by Theorem \ref{the-1}, with probability $1-\nu$, $\mathcal{C}_{f-f^*}(p_{\alpha_{j_0}},\mathbf{s}_{j_0})\leq \lfloor\frac{3}{8}T_1\rfloor$.
If $j_0$ satisfies $\mathcal{C}_{f-f^*}(p_{\alpha_{j_0}},\mathbf{s}_{j_0})\leq \lfloor\frac{3}{8}T_1\rfloor$, then by Lemma \ref{lm-5}, there are at most $\lfloor\frac{3}{8}T_1\rfloor$ terms in $f-f^*$ but not in $h$.
Since the terms of $h$ which are not in $f-f^*$ come from at least three terms in $f-f^*$, then there are at most $\frac13\lfloor\frac{3}{8}T_1\rfloor$ terms of $f^*$ not in  $f$. So $\#(f-f^*-h)\leq \lfloor\frac{3}{8}T_1\rfloor+\frac13\lfloor\frac{3}{8}T_1\rfloor\leq \frac12T_1$.
So we have $\#(f-f^*-h)\leq \lfloor \frac12T_1\rfloor$.
The first part is proved.

Now we analyse the complexity.
In Step 1, use the sieve of Eratosthenes \cite[p.500, Them.18.10]{9},  the cost of finding the $N$ primes bigger than $32(T_1-1)$ is  $O^\thicksim(T_1\log D)$ bit operations.

In Step 2, since probabilistic machines flip coins to decide binary digits, each of these random choices can be simulated with a machine with complexity $O(n\log (T_1\log D))$. So the complexity of Step 2 is $O(n\log^2T_1+n\log T_1 \log\frac{1}{\nu}+n\log T_1\log\log D+n\log\log D\log\frac{1}{\nu})$ bit operations.

In Step 3, since $p_{\alpha_i}$ is $O^\thicksim(T_1\log D)$, the degree of $f(x^{\mathbf{s}_i})$ is $O^\thicksim(\|\mathbf{s}_i\|_\infty D)=O^\thicksim(T_1D)$.
So in Step 3, we query $O(\log T_1+\log \frac{1}{\nu})$ polynomials of degree $O^\thicksim(T_1D)$. In order to obtain $f^*(x^{\mathbf{s}_i})$, it needs $O(lnT)$ ring operations and $O^\thicksim(lnT\log D)$ bit operations. In order to obtain $f_i$, it needs $O(lT)$ ring operations in $\RB$ and $O^\thicksim(lT\log D)$ bit operations. In order to obtain the $f_i^{\modp}$, it needs $O^\thicksim(lT_1\log D)$ bit operations and $O(lT_1)$ ring operations.
So it still needs $O^\thicksim(nT\log \frac{1}{\nu}+T_1\log \frac{1}{\nu})$ ring operations and $O^\thicksim(nT\log D\log \frac{1}{\nu}+T_1\log D\log \frac{1}{\nu})$ bit operations.

In Step 4, we find the integer $j_0$. Since $\#f^{\modp}_i\leq T_1$, it needs at most $O^\thicksim(T_1\log \frac{T_1}{\nu})$ bit operations to compute all $\#f^{\modp}_i,i=1,2,\dots,l$.
Find $j_0$ needs $O^\thicksim(l\log T_1)$ bit operations. So the bit complexity of Step 4 is $O^\thicksim(T_1\log \frac{1}{\nu})$.

In Step 5, since the degree of $f(x^{\mathbf{s}_{j_0}+p_{\alpha_{j_0}}\mathbf{I}_k})$ is $O^\thicksim(T_1D)$, it queries $O(n)$ polynomials of degrees $O^\thicksim(T_1D)$. By Lemma \ref{lm-m1}, it needs
$O^\thicksim(nT\log D)$ bit operations and $O(nT)$ arithmetic operations in $\RB$ to obtain $\{f^*_1,f^*_2,\dots,f_n^*\}$.

%


In Step 6, by Theorem \ref{the-3}, the complexity is $O(nT_1)$ ring operations in $\RB$ and $O^\thicksim(nT_1\log D)$ bit operations.
Since $T\geq T_1$, the lemma is proved.\qed

We now give the complete interpolation algorithm.

\begin{alg}[MulPolySI]\label{alg-m3}
\end{alg}

{\noindent\bf Input:}
A Black-box procedure $\mathcal{B}_f$ that computes $f\in\RB[\X]$,
$T\geq \# f$,
 $D\geq \deg(f)$,
and  $\mu\in(0,1)$.

{\noindent\bf Output:}
Return $f$ with probability $\geq1-\mu$,  or
failure.
\begin{description}
\item[Step 1:] Let $h=0,T_1=T,\nu=\frac{\mu}{\lceil\log_2 T\rceil+1}$.

\item[Step 2:] While $T_1>0$ do

\begin{description}
\item[b:] Let $g=\mathbf{HalfPoly}(\mathcal{B}_f,h,T,T_1,D,\nu)$. If $g=failure$, then return failure.

\item[c:] Let $h=h+g$, $T_1=\lfloor\frac{T_1}{2}\rfloor$.
\end{description}

\item[Step 3:] Return $h$.

\end{description}

\begin{theorem}\label{them-5}
Algorithm \ref{alg-m3} computes $f$ with probability $\ge 1-\mu$.
The algorithm needs

$\bullet$ $O(n\log T+\log^2 T+\log T\log \frac{1}{\mu})$ interpolations of univariate polynomials with degree $O^\thicksim(TD)$ and sparseness $\le T$.

$\bullet$ $O^\thicksim(nT\log \frac{1}{\mu})$ additional ring operations  and $O^\thicksim(nT$ $\log D\log\frac{1}{\mu})$ additional bit operations.
\end{theorem}
\proof
In  $\mathbf{b}$ of Step 2,
since $\#(f-h)\le  T_1$, by Lemma \ref{them-3}, $\#(f-h-g)\leq \lfloor \frac{T_1}{2}\rfloor$ with probability $\ge 1-\nu$.
Then, Step 2 will run at most $k= \lceil\log_2 T\rceil+1$ times and return the correct $f$ with  probability $\geq(1-\nu)^k\geq 1-\mu$. The first part is proved.

Now we analyse the complexity. It is easy to see that the complexity is dominated by  Step 2.
In Step 2, we call at most $O(\log T)$ times Algorithm \ref{alg-2}. Since the terms and degrees of $f-h$ are respectively bounded by $T$ and $D$, by Theorem \ref{them-3}, it needs $O(n\log T+\log^2 T+\log T\log \frac{1}{\nu})$ queries of degree $O^\thicksim(TD)$,
$O^\thicksim(nT\log \frac{1}{\nu})$ additional ring operations  and $O^\thicksim(nT\log D\log \frac{1}{\nu})$ additional bit operations.
Since $\nu=\frac{\mu}{\lceil\log_2 T\rceil+1}$, we have proved the theorem.\qed

\begin{cor}\label{rem-1}
Set $\mu=1/4$. Then
Algorithm \ref{alg-m3} computes $f$ with probability at lest $\frac34$. The algorithm needs

$\bullet$ $O(n\log T+\log^2 T)$ queries of univariate polynomials with degree $O^\thicksim(TD)$ and sparseness $\le T$.

$\bullet$ $O^\thicksim(nT)$ additional ring operations  and $O^\thicksim(nT\log D)$ additional bit operations.
\end{cor}


\section{Sparse interpolation over finite fields}
In this section, we give a sparse interpolation algorithm for black-box multivariate polynomials over general finite fields.
We first  give an univariate Ben-or and Tiwari algorithm over finite fields and then combine with Algorithm \ref{alg-m3}  to give a multivariate interpolation algorithm.

\subsection{The Ben-Or and Tiwari sparse interpolation algorithm}
Following \cite{7}, we give a brief introduction to
the multivariate Ben-Or and Tiwari sparse interpolation algorithm over  $\C$.

Let $f(x_1, \dots, x_n)=c_1m_1+\cdots+c_t m_t\in\C[\X]$
be the polynomial to be interpolated,  where $m_i=x_1^{e_{i,
1}}\dots x_n^{e_{i, n}}$ are distinct monomials,  $c_i$ are non-zero
coefficients,  and $t=\#f$ is the number of terms in $f$.
We assume that $f$ is a {\em black-box},  which means,  for
$\forall$ $(q_1, \dots, q_n)\in \C^n$,  we can obtain the value
$f(q_1, \dots, q_n)$. Note that $c_i,  m_i, t$ are not known. In
order to determine $f$ uniquely,   the algorithm needs as input an
upper bound $\tau+1\geq t$ on the number of terms in $f$.

The algorithm proceeds in two stages. The monomials $m_i$ are determined first using an auxiliary polynomial $\zeta(z)$. Once the $m_i$ are known,  the coefficients $c_i$ can be obtained easily.

We first determine $m_i$. Let $v_i=p_1^{e_{i, 1}}\dots p_n^{e_{i,
n}}$ denote the value of the monomial $m_i$ at $(p_1, \ldots, p_n)$,
where $p_i$ is the $i$-th prime number. Clearly,  different
monomials evaluate to different values under this evaluation. Let
$a_0, a_1, \dots, a_{2\tau+1}$ be the values of $f$ at the
$2(\tau+1)$ points $\p_i=(p_1^i, \ldots, p_n^i),
i=0,1,\ldots,2\tau+1$,  that is,  $a_i=\sum_{j=1}^{t} c_j v_j^i$.

The auxiliary polynomial $\zeta(z)$ is defined as follows.
 \begin{eqnarray}
   \zeta(z)=\prod_{i=1}^{t} (z-v_i) =z^t+\zeta_{t-1}z^{t-1}+\dots+\zeta_1 z+\zeta_0.
\end{eqnarray}

Consider the sum
$
\sum_{i=1}^{t} c_i v_i^j \zeta(v_i)=\sum_{k=0}^{t-1} \zeta_k(c_1 v_1^{k+j}+c_2 v_2^{k+j}+\dots+c_t v_t^{k+j})+(c_1 v_1^{t+j}+c_2 v_2^{t+j}+\dots+c_t v_t^{t+j})
=a_j \zeta_0+a_{j+1} \zeta_1+\dots+a_{j+t-1}\zeta_{t-1}+a_{j+t}$
for $j=0, \dots, t-1$.
Since $\zeta(v_i)=0$,  for $0\leq j \leq t-1$, we have
\begin{eqnarray}\label{eq-toeplitz}
 a_j \zeta_0+a_{j+1} \zeta_1+\dots+a_{j+t-1}\zeta_{t-1}+a_{j+t}=0.
\end{eqnarray}
This is a Toeplitz system $T_{t-1, t-1} \hat{\zeta}_{t-1}=\hat{t}_{2t-1, t-1}$  where

\centerline{
$T_{u, v}=\left(\begin{array}{cccc}
a_u&a_{u+1}&\cdots&a_{u+v}\\
a_{u-1}&a_u&\cdots&a_{u+v-1}\\
\vdots&\vdots&\ddots&\vdots\\
a_{u-v}&a_{u-v+1}&\cdots&a_u\\
\end{array}\right)$}
%
\noindent
 $\hat{\zeta}_v=(\zeta_0,\zeta_1,\ldots,\zeta_v)^\tau$,
 $\hat{t}_{u, v}=-(a_u,a_{u-1},\ldots,a_{u-v})^\tau$.
This system is non-singular as can be seen from the factorization.
{\small \begin{eqnarray}\label{eq-vand2}
T_{t-1,t-1}&=&\left(\begin{array}{cccc}
1&1&\cdots&1\\
v_1&v_2&\cdots&v_t\\
\vdots&\vdots&\ddots&\vdots\\
v_1^{t-1}&v_2^{t-1}&\cdots&v_t^{t-1}\\
\end{array}\right)
\left(
\begin{array}{cccc}
c_1&0&\cdots&0 \\
0&c_2&\cdots&0 \\
\vdots&\vdots&\ddots&\vdots \\
0&0&\cdots&c_t\\
\end{array}
\right)\left(
\begin{array}{cccc}
1&v_1&\cdots&v_1^{t-1} \\
1&v_2&\cdots&v_2^{t-1}\\
\vdots&\vdots&\ddots&\vdots \\
1&v_t&\cdots&v_t^{t-1}\\
\end{array}
\right)
\end{eqnarray}}

Since the $v_i$ are distinct, the two Vandermonde matrices are nonsingular and as no $c_i$ is zero, the diagonal matrix is nonsingular, too. If the input value of the upper bound $\tau+1$ is greater than $t$, then the coefficients $c_k$, for $k>t$, can be regarded as zero and the resulting system $T_{\tau,\tau}$ would be singular.

\begin{lemma}[\cite{7}]\label{eq-lm1}
If $t$ is the exact number of terms in $f$,  then

$a)$ $T_{i, t-1}$is non-singular for all $i\geq t-1$.

$b)$ $T_{i, t+j}$ is singular for all $i\geq t-1, j\geq0$.
\end{lemma}
%

By Lemma \ref{eq-lm1},  when considering $2\tau+2$ values $a_0,
\ldots, a_{2\tau+1}$ of $f$,  the coefficients of $\zeta(z)$ can be
uniquely recovered from the system
$T_{\tau, \tau} \hat{\zeta}_{\tau}=\hat{t}_{2\tau+1, \tau}$.
By finding the roots $v_i=p_1^{e_{i, 1}}\dots p_n^{e_{i, n}}$ of $\zeta(z)$,  the monomials $m_i$ can be recovered.

By choosing the first $t$ evaluations $a_0, \ldots, a_{t-1}$ of $f$,  we obtain the following transposed Vandermonde system $A\hat{c}=\hat{a}$ for the coefficients of $f$,  where
{\small
\begin{equation}\label{eq-vand}
A=\left(\begin{array}{cccc}
1&1&\cdots&1\\
v_1&v_2&\cdots&v_t\\
\vdots&\vdots&\ddots&\vdots\\
v_1^{t-1}&v_2^{t-1}&\cdots&v_t^{t-1}\\
\end{array}\right),
\hat{c}=\left(
\begin{array}{r}
c_1 \\
c_2 \\
\vdots \\
c_t\\
\end{array}
\right),
\hat{a}=\left(
\begin{array}{c}
a_0 \\
a_1 \\
\vdots \\
a_{t-1}\\
\end{array}
\right)
\end{equation}}

The deterministic Ben-or and Tiwari's algorithm over $\Z$ needs $O(T)$ evaluations of $f$ plus  $O(nT^2d)$
$\Z$-operations and the height of the data is $Td$ \cite{7},
where $d= \deg f$.
%

If the coefficients of the polynomials are from a finite field,
then it is difficult to find the exponents from $v_i=p_1^{e_{i, 1}}\dots p_n^{e_{i, n}}$, which is a multi-variate discrete logarithm problem.

\subsection{ Univariate Ben-or and Tiwari algorithm over finite field}
In this section, we give a modified univariate Ben-or and Tiwari algorithm over the finite field $\F_q$.
%
Assume
$f(x)=\sum_{i=1}^tc_im_i\in\F_q[x],D\geq\deg(f)$.
Since  $f(x)$ is univariate, $\#f\le D$.
We consider two cases:  $q>D$ or $q\le D$.

First, consider the case $q>D$.
%
Let $\omega$ be  a primitive element of $\F_q$. Assume $m_i=x^{d_i}$ and denote $v_i=\omega^{d_i}$.
Let $a_i=\sum_{j=1}^{t} c_j v_j^i, i=0,1,\ldots,2\tau+1$.
$T_{t-1,t-1}$ still can be factored as \bref{eq-vand2}. Since $\omega$ is a primitive element of $\F_q$ and $q>D$, $v_i\neq v_j$ when $i\neq j$. So the two Vandermonde matrices in \bref{eq-vand2} are nonsingular and  Lemma \ref{eq-lm1} is still correct.
Now we can give the  algorithm.
\begin{alg}[UniBoTFq]\label{alg-bt2}
\end{alg}

{\noindent\bf Input:} A black-box procedure $B_f$ to compute $f(x)\in\F_q[x]$, $\tau+1\ge \#f$, and $D\geq\deg f$.

{\noindent\bf Output:} The polynomial $f=\sum_{i=1}^t c_i m_i$.
\begin{description}
\item[Step 1:] Let $\omega$ be a primitive element of $\F_q$. Evaluate $f$ at the $2(\tau+1)$ points $\omega^i$, $i=0, \dots, 2\tau+1$.
 Let $a_i,i=0, \dots, 2\tau+1$ be the corresponding values.

\item[Step 2:]  Solve the Toeplitz system $T_{\tau, \tau}\hat{\zeta}_\tau=\hat{t}_{2\tau+1, \tau}$ (or the largest non-singular subsystem $T_{j, 2\tau-j}$ $ \widehat{\zeta}_{2\tau-j}=\hat{t}_{2\tau+1, 2\tau-j}$ of $T_{\tau, \tau}$,  where $j$ is the smallest positive integer that makes $T_{j, 2\tau-j}$ non-singular) to obtain the polynomial $\zeta(z)=\sum_{i=0}^t \zeta_i z^i$.


\item[Step 3:]  Find the monomial set $M$ of $f$. $M=\emptyset$.
For $i=0,1,\dots,D$,
compute $\omega^i$ and if $\zeta(\omega^i)=0$ then let $M=\{x^i\}\cup M$.

\item[Step 4:] Find the coefficients $c_i$ by solving the transposed Vandermonde system $A\hat{c}=\hat{a}$ in \bref{eq-vand}.
\end{description}

\begin{lemma}\label{lm-4}
If $q>D$, Algorithm \ref{alg-bt2} is correct and  it needs $2(\tau+1)$ evaluations of $f$ plus $O^\thicksim(D\log q)$ bit operations.
\end{lemma}
\proof
The correctness comes from Lemma \ref{eq-lm1}. Now we analyse the complexity.
Due to the fast integer and polynomial multiplication algorithms \cite[p.232]{9}, one can perform an arithmetic operation in $\F_{q}$ in $O^\thicksim(q)$ bit operations.
In Step 1, it needs $O(\tau\log q)$ bit operations to obtain $\omega^i$ and $a_i$, $i=1,2,\dots,2\tau+1$. In Step 2, it needs $O(M(\tau)\log \tau\log q)$ bit operations, where $M(\tau)=\tau\log(\tau)\log\log(\tau)$ \cite{7}.

In Step 3, computing $\omega^i,i=0,1,\dots,D$ needs $O(D\log q)$ bit operations. Then we evaluate $\zeta(\omega^i),i=0,1,\dots,D$, by fast multi-point evaluation method \cite[p.298.Them.10.6]{9}, which needs $O(\frac{D}{T}M(T)\log T\log q)=O^\thicksim(D\log T\log q)$ bit operations, where  $T=\tau+1$. 

In Step 4, it needs $O(M(t)\log t\log q)$ bit operations \cite{7}.
So the complexity of the total algorithm is  $O^\thicksim(D\log T\log q+T\log q)
= O^\thicksim(D\log q)$  bit operations, since $\#f \le D$.\qed

\vskip10pt
Second, consider the case $q<D$.
%
We need evaluate the polynomial in an extended field of $\F_q$. We extends $\F_q$ into $\F_{q^m}$ such that $q^m\geq D+1$, where $m= \lceil\frac{\log (D+1)}{\log q}\rceil$.
Due to the fast integer and polynomial multiplication algorithms \cite[p.232]{9}, one can perform an arithmetic operation in $\F_{q^m}$ in $O^\thicksim(m\log q)
= O^\thicksim(\log D)$ bit operations, since $m= \lceil\frac{\log (D+1)}{\log q}\rceil$.
%

Now we can extend Algorithm \ref{alg-bt2} into the case $q\le D$. The only change is to replace the primitive element of $\F_{q}$ by a primitive element of $\F_{q^m}$ in Step 1.
Similar to the proof of Lemma \ref{lm-4}, the complexity of the algorithm is $O^\thicksim(D\log Tm \log q +Tm\log q)$, which is $O^\thicksim(D\log T+T\log D)
=O^\thicksim(D\log D) = O^\thicksim(D)$ bit operations.
We thus have
\begin{lemma}\label{cor-4}
If $q\le D$, Algorithm \ref{alg-bt2}  needs $2(\tau+1)$ evaluations of $f$ plus $O^\thicksim(D)$ bit operations.
\end{lemma}
%

Following Lemmas \ref{lm-4} and \ref{cor-4}, we have
\begin{theorem}\label{the-4}
Let $f$ be a black-box univariate polynomial in $\F_q[x]$
with $T\ge \#f$ and $D\ge \deg f$.
We can compute $f$ with $O(T)$ evaluations of $f$ plus $O^\thicksim(D\log q)$ bit operations.
\end{theorem}

\begin{remark}\label{rem-dl1}
In Step 3 of Algorithm \ref{alg-bt2}, we may follow the original Ben-or and Tiwari algorithm to find the exponents.
%
First, find the roots $v_i$ of $\zeta(z)=0$, which costs
$O^\thicksim(t\log^2 q)$ bit operations \cite[p.368]{9} for $t=\#f$.
Second, solve the discrete logarithm problem $v_i = \omega^{e_i}$ to find the exponents $e_i$, which costs $O^\thicksim(\sqrt{D}\log q)$ bit operations \cite{dislog}.
Therefore, the total complexity of the algorithm is
$O^\thicksim(T\log^2 q+T\sqrt{D}\log q)$ bit operations plus $O(T)$ evaluations.
\end{remark}

\subsection{Multivariate polynomial interpolation over finite fields}
Combing the reduction algorithm given in Section 2 and the univariate
interpolation given in Section 3.2, we give a multivariate interpolation algorithm over finite fields.

%
\begin{theorem}\label{th-ff}
Let $f\in\F_q[\X]$ be a black-box polynomial.
Given $T\ge \#f$ and $D\ge \deg(f)$,  with probability greater than $\frac{3}{4}$, one can find   $f$ using $O^\thicksim(nTD\log q)$ bit operations plus $O^\thicksim(nT)$ evaluations of $f$.
\end{theorem}
\proof
We use the Algorithm \ref{alg-m3} to compute $f$ and use
Algorithm \ref{alg-bt2} for  univariate polynomial
interpolation in Step 2 and Step 5 of \ref{alg-2}.

The complexity consists of two parts.
By Corollary \ref{rem-1}, we needs $O(n\log T+\log^2 T)$ queries of univariate polynomials with degree $O^\thicksim(TD)$ and sparseness $\le T$. Then by Theorem \ref{the-4}, we need $O^\thicksim((n\log T+\log^2 T)T) =
O^\thicksim(nT)$ evaluations of $f$ and
$O^\thicksim((n\log T+\log^2 T)(TD\log q)) =
O^\thicksim(nTD\log q)$ bit operations to query these univariate polynomials.

By Corollary \ref{rem-1}, we needs additional
$O^\thicksim(nT)$ operations in $\F_q$ if $q > D$ (or in $\F_{q^m}$ if $q< D$ for $m= \lceil\frac{\log (D+1)}{\log q}\rceil$)
and $O^\thicksim(nT\log q)$ bit operations.
$O^\thicksim(nT)$ operations in $\F_q$ costs $O^\thicksim(nT\log q)$ bit operations.
$O^\thicksim(nT)$ operations in $\F_{q^m}$ costs $O^\thicksim(nTD)$ bit operations.
Therefore, the query of $f$ is the dominate step
and the bit complexity of the algorithm is $O^\thicksim(nTD\log q)$.\qed

\begin{remark}\label{rem-dl2}
If using the original Ben-or and Tiwari algorithm mentioned in Remark \ref{rem-dl1} to interpolation the univariate polynomials,
the total complexity of our algorithm is
$O^\thicksim(nT^{1.5}\sqrt{D}\log q+nT\log^2 q)$ bit operations.
\end{remark}

\begin{remark}
Let $f\in\F_q[\X]$ be a black-box polynomial.
If quantum algorithms can be used, the quantum complexity
of finding $f$ is $O^\thicksim(nT \log^2 q)$ plus $O^\thicksim(nT)$ evaluations of $f$ and $O^\thicksim(nT)$
black-box evaluations for solving the discrete logarithm problem.

We need to change step 3 of Algorithm \ref{alg-bt2} as follows:

(1) Find the roots $v_i$ of $\zeta(z)$, which costs an expected $O^\thicksim(T$ $ \log^2 q\})$ bit operations \cite[p.368]{9}.

(2) Solve the discrete logarithm problem $v_i=\omega^{e_i}\modp\, q$ to find ${e_i}$ using Shor's quantum algorithm, which costs $O^\thicksim(T$ $\max\{\log^2 D,$ $ \log^2 q\})$ plus $T$
black-box evaluations \cite[p.238]{qc}.

Since $D\le n(q-1)$,  by Corollary \ref{rem-1},
the total complexity is  $O^\thicksim(nT \max\{\log^2 D,$ $ \log^2 q\})
= O^\thicksim(nT \log^2 q)$.
%
\end{remark}

\section{Experimental results}

In this section, practical performances of the interpolation
algorithm over finite fields given in Remark \ref{rem-dl2} will be reported.
The algorithm uses Algorithm \ref{alg-m3} to reduce multivariate
interpolation to univariate interpolation and uses
Algorithm \ref{alg-bt2} for univariate polynomial
interpolation.
In Algorithm \ref{alg-bt2}, we use the Berlekamp-Massey algorithm to solve the Toeplitz systems, use the command $Roots$ in Maple to find the roots, and use the command $mlog$ in Maple to solve the discrete logarithm problem.

The data are collected on a desktop with Windows system,
3.60GHz Core $i7$-$4790$ CPU, and 8GB RAM memory.
The implementations in Maple can be found in
\begin{verbatim}
http://www.mmrc.iss.ac.cn/~xgao/software/rkron.zip
\end{verbatim}

We randomly construct five polynomials over the finite field $\F_q$, then regard them as black-box polynomials
and reconstruct them with the algorithm. The actual size and degree of the polynomials are used as the term bound and degree bound, respectively. The average times are collected. In our testing, we fix $q=30000000001$ and use the primitive element $29$ of $\F_q$.

\begin{figure}[ht]
\begin{minipage}[t]{0.9\linewidth}
\centering
\includegraphics[scale=0.40]{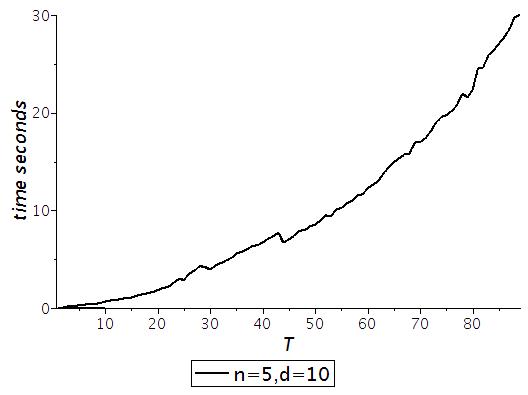}
\caption[KST.jpg]{Average   times with varying $T$} \label{fig1}
\end{minipage}\quad
\end{figure}

\begin{figure}[ht]
\begin{minipage}[t]{0.95\linewidth}
\centering
\includegraphics[scale=0.40]{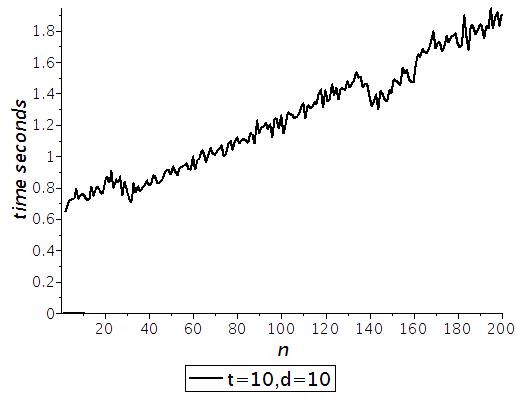}
\caption[KSn.jpg]{Average   times with varying $n$} \label{fig2}
\end{minipage}\quad
\end{figure}

\begin{figure}[ht]
\begin{minipage}[t]{0.95\linewidth}
\centering
\includegraphics[scale=0.40]{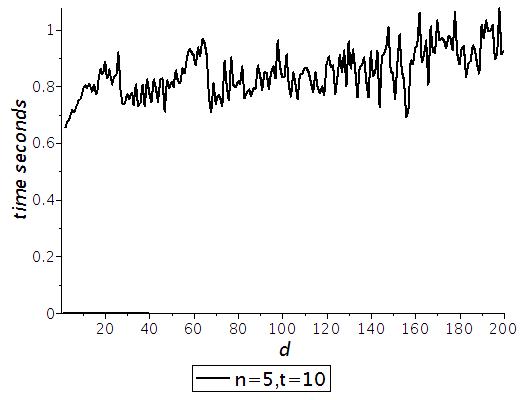}
\caption{Average   times with varying $d$ }\label{fig3}
\end{minipage}
\end{figure}

The results are shown in Figures \ref{fig1}, \ref{fig2}, \ref{fig3}.
In each figure, two of the parameters $n,T,D$ are fixed and one of them is variant.
%
%
%
%
These data are basically in accordance with the complexity
$O^\thicksim(nT^{1.5}\sqrt{D}\log q+nT\log^2 q)$ of the algorithm.

\section{Conclusion}
In this paper, we revisit the approach of reducing
the black-box multivariate polynomial interpolation to that of
the univariate polynomials by randomized Kronecker Substitution
and give an algorithms with better complexities in most cases.
%
%
The algorithm consists of two main ingredients.
First, we give a reduction method which reduces the interpolation
of multivariate polynomials to that of univariate ones, which
need to interpolate $O(n\log T+\log ^2 T)$  univariate polynomials of degree $O^\thicksim(TD)$
and an extra $O(nT\log D)$ bit operations.
Second, we give a modified Ben-or and Tiwari algorithm over
the finite file $\F_q$ for a univariate polynomial of degree $d$
and term $t$, which costs $O^\thicksim(d\log t\log q+t\log q)$
bit operations.
Combing the two ingredients, we obtain a multivariate interpolation algorithm over the finite field $\F_q$ which needs $O^\thicksim(nT)$ evaluations of the black-box plus $O^\thicksim(nTD\log q)$ bit operations.

\end{document}